\documentclass[conference,letterpaper]{IEEEtran}
\usepackage{cite}
\usepackage{graphicx}
\usepackage[cmex10]{amsmath}
\usepackage{fixltx2e}
\usepackage{amssymb}
\usepackage{algpseudocode}

\newsavebox{\ieeealgbox}
\newenvironment{boxedalgorithmic}
{\begin{lrbox}{\ieeealgbox}
		\begin{minipage}{\dimexpr\columnwidth-2\fboxsep-2\fboxrule}
			\begin{algorithmic}}
			{\end{algorithmic}
		\end{minipage}
	\end{lrbox}\noindent\fbox{\usebox{\ieeealgbox}}
	}

\begin{document}

\title{Design of LDPC Code Ensembles\\with Fast Convergence Properties}

\author{\IEEEauthorblockN{Ian P. Mulholland\IEEEauthorrefmark{1},
Enrico Paolini\IEEEauthorrefmark{2} and
Mark F. Flanagan\IEEEauthorrefmark{1}}
\IEEEauthorblockA{\IEEEauthorrefmark{1}School of Electrical, Electronic and Communications Engineering, University College Dublin, Dublin, Ireland\\ Email: \{mulholland, mark.flanagan\}@ieee.org}
\IEEEauthorblockA{\IEEEauthorrefmark{2}Department of Electrical, Electronic, and Information Engineering ``G. Marconi'', University of Bologna, Cesena (FC), Italy\\
Email: e.paolini@unibo.it}}

\maketitle

\begin{abstract}
The design of low-density parity-check (LDPC) code ensembles optimized for a finite number of decoder iterations is investigated. Our approach employs EXIT chart analysis and differential evolution to design such ensembles for the binary erasure channel and additive white Gaussian noise channel. The error rates of codes optimized for various numbers of decoder iterations are compared and it is seen that in the cases considered, the best performance for a given number of decoder iterations is achieved by codes which are optimized for this particular number. The design of generalized LDPC (GLDPC) codes is also considered, showing that these structures can offer better performance than LDPC codes for low-iteration-number designs. Finally, it is illustrated that LDPC codes which are optimized for a small number of iterations exhibit significant deviations in terms of degree distribution and weight enumerators with respect to LDPC codes returned by more conventional design tools.
\end{abstract}

\IEEEpeerreviewmaketitle

\section{Introduction}

Since their rediscovery in the $1990$s \cite{MacKay1999Gooderrorcorrecting}, the design of LDPC code ensembles \cite{Gallager1963LowDensityParity} has often focused primarily on their impressive capacity-approaching capabilities. The dominant approaches to design for optimizing the decoding threshold of LDPC codes are density evolution \cite{Richardson2001Designcapacityapproaching} and extrinsic information transfer (EXIT) chart \cite{TenBrink2001Convergencebehavioriteratively} analysis.

Whereas traditionally iteratively decodable codes such as LDPC codes have been designed under the assumption of an infinite number of decoder iterations, in practice it is preferable that the maximum number of iterations considered in the design be the one targeted by the specific application considered. Traditional code design methods based on threshold optimization necessarily fail to capture the effect of the maximum number of iterations in practical scenarios, and the code designs so produced may not offer the best - or even good - performance for such scenarios.

The number of decoder iterations required for successful iterative decoding has recently attracted interest.  The design of non-systematic irregular repeat-accumulate codes and LDPC codes optimized for a finite number of iterations over the binary erasure channel (BEC) was investigated in \cite{Grant2012Iterationconstraineddesign} and \cite{Jamali2014efficientcomplexityoptimizing}, respectively, while lower bounds on the number of iterations required for decoding of turbo-like ensembles were found in \cite{Sason2009BoundsNumberIterations}.

In this paper we propose an ensemble design strategy which takes the maximum number of iterations as an input. This ensemble design strategy, which is based on a modified analysis using EXIT charts combined with optimization using differential evolution, is used to design LDPC code ensembles which are optimized for a finite number of iterations over the BEC and binary input additive white Gaussian noise (BI-AWGN) channel. It is seen that for both channels, when the maximum number of decoder iterations, $i_{\mathrm{max}}$, is constrained, LDPC codes which are optimized for $i_{\mathrm{max}}$ iterations outperform those which are optimized for an infinite number of iterations. Our strategy, in which the problem of fast convergence is approached from a code design perspective, contrasts with the strategies adopted in most of the literature, in which fast convergence is tackled from a decoding algorithm point of view \cite{Zhang2005Shufflediterativedecoding, Hocevar2004reducedcomplexitydecoder, Xia2013newstoppingcriterion}. Instead, we are interested in ensembles such that a code in the ensemble converges quickly with high probability.

When a small number of iterations is targeted, our approach returns ensembles with a number of different features with respect to traditional approaches. It is well known that in order for an LDPC code ensemble to approach capacity, a large fraction of degree-$2$ variable nodes (VNs), is required. In contrast, in our iteration-constrained ensembles this fraction is very small when the maximum number of iterations is small. Also, in contrast to ensembles which are optimized for a large or unlimited number of iterations, these ensembles have good behavior with respect to the growth rate of the weight distribution and with respect to the typical minimum distance. We also consider the optimization of generalized LDPC (GLDPC) code ensembles, in which the single parity-check (SPC) codes at the check nodes (CNs) of an LDPC code ensemble are replaced with generic linear block codes. Imposing constraints on the number of decoder iterations for the BEC, we show that the presence of these generalized CNs may be beneficial to reduce the number of message exchanges to achieve a target performance.

\section{Notation}
\label{section_preliminaries} 

We define a GLDPC code ensemble as follows. The block length of a code in the ensemble is denoted by $N$. There are $n_c$ different CN types $t\in I_c = \{1,2,\dots,n_c\}$. For each CN type $t\in I_c$ we associate a local linear block code with length $s_t$, code rate $R_t$, and minimum distance $r_t$. Each local code may be a single parity-check code of length $s_t$, or a more general linear block code. In the special case where all local codes are SPC codes, we have a traditional LDPC ensemble. The polynomial $\rho(x)$ is defined by $\rho(x) = \sum_{t\in I_c}\rho_t x^{s_t-1}$, where for each $t \in I_c$, $\rho_t$ denotes the fraction of edges connected to CNs of type $t$. Similarly, we define $\lambda(x) = \sum_{d\in I_v} \lambda_d x^{d-1}$, where for each $d \in I_v = \{2,3,\dots,d_v\}$, $\lambda_d$ denotes the fraction of edges connected to VNs of degree $d$. The design rate of the ensemble, is given by

\begin{equation*}
R = 1 - \frac{\sum_{t\in I_c}\rho_t(1-R_t)}{\sum_{d\in I_v}\lambda_d/d}.
\end{equation*}

For ensembles with a positive fraction of CNs having minimum distance $2$, we define

\begin{equation*}
C = 2\sum_{t:r_t = 2}\frac{\rho_tA_2^{(t)}}{s_t},
\end{equation*}

\noindent where $A_2^{(t)}$ denotes the number of weight-$2$ codewords for CNs of type $t$. Moreover, $E = N/{\int\lambda}$ denotes the number of edges in the Tanner graph. The number of CNs of type $t \in I_c$ is given by $E\rho_t/s_t$.

\section{Ensemble Design Strategy}
\label{section_design_strategy}

In EXIT chart analysis, the \textit{a-priori} information at a node is mapped to the extrinsic output information of that node by an EXIT function. On the VN side of the Tanner graph, the average \textit{a-priori} and extrinsic information are denoted by $I_{A,V}$ and $I_{E,V}$, respectively. Similarly, $I_{A,C}$ and $I_{E,C}$ denote the average \textit{a-priori} and extrinsic information, respectively, on the CN side of the graph. By plotting the VN-side and CN-side EXIT functions against one another on an EXIT chart for a given value of the channel parameter (for example, the erasure probability $\varepsilon$ and the signal-to-noise ratio $E_b/N_0$ for the BEC and BI-AWGN channel, respectively), it is possible to visualize the passing of extrinsic information between the two sides  as a ``decoding path'' moving between the two EXIT curves iteration by iteration. The \emph{iterative decoding threshold} for an ensemble is the most unfavorable channel parameter for which the EXIT functions do not intersect.

While the use of EXIT charts in this way can lead to impressive thresholds when designing ensembles, such thresholds may be of limited usefulness in practical scenarios with severe constraints on the maximum number of iterations. In order to design ensembles for such scenarios, the traditional EXIT analysis should be modified. We remove the condition that the EXIT functions must not intersect and instead require that after $i_{\mathrm{max}}$ decoder iterations, the output extrinsic information must exceed some value $\xi$. We define the \emph{iteration-constrained threshold} of an ensemble as the worst value of the channel parameter (e.g., the highest value of $\varepsilon$ or lowest value of $E_b/N_0$ for the BEC or BI-AWGN channel, respectively) for which this requirement is satisfied. One consequence of this new condition is that even when the EXIT curves intersect, we may consider a channel parameter to be achievable, provided that the intersection occurs at some point where $I_{A,V} > \xi$.

Differential evolution is an optimization algorithm often used to find, for some design rate $R$, a degree distribution pair (DDP) $(\lambda, \rho)$ which gives rise to a good threshold \cite{Shokrollahi2000Designefficienterasure}. An initial population containing $N_p$ members is generated, where each member is a length-$D$ vector which expresses the DDP of an ensemble of design rate $R$. A trial vector is then created for each member of the population by combining a subset of randomly chosen vectors from the population. If this competing vector offers a better threshold than its corresponding population member, the member is replaced. This process is repeated until a stopping criterion has been fulfilled. The steps involved in differential evolution, as used in this paper, are given in Fig. \ref{algorithm_DE}. The output of the algorithm is the best member of the population after the stopping criterion has been reached, i.e., the vector in the final population with the best threshold. It is important to observe that each time that a new vector is created (i.e., in steps \ref{algorithm_DE_generate_mutant_vector} and \ref{algorithm_DE_generate_trial_vector}), it is unlikely that this vector will  still meet the target design rate $R$ and also satisfy $\sum_d \lambda_d = 1$ and $\sum_t \rho_t = 1$. Therefore, during the creation of each trial vector $\mathbf{u}_i^l$ in step \ref{algorithm_DE_generate_trial_vector}, it is necessary to adjust three of the $D$ elements in the vector in order to produce a vector which satisfies all three constraints. In the event that some vector elements which are not in $[0,1]$ are obtained, these vectors are discarded and new ones are generated.

By using our iteration-constrained EXIT chart analysis to examine the decoding performance of ensembles generated in each iteration of differential evolution, it is possible to design ensembles optimized for a given $i_{\mathrm{max}}$.

\begin{figure}[hb]
\begin{boxedalgorithmic}[1]
	\State Generate a random starting population $\{\mathbf{x}_1^0,\dots,\mathbf{x}_{N_p}^0\}$ where each $\mathbf{x}_i^0$ is some degree distribution pair $(\lambda,\rho)$ with the target design rate $R$. Let $l=0$.
	\State For each $i\in \{1,\dots,N_p\}$, generate a mutant vector $\mathbf{v}_i^l$ where $\mathbf{v}_i^l = \mathbf{x}_{r_1}^l + F(\mathbf{x}_{r_2}^l-\mathbf{x}_{r_3}^l)$ where $r_1,r_2,r_3$ are picked uniformly at random without replacement from $\{1,\dots,N_p\}\backslash \{i\}$ and $F$ is a constant, usually between $0$ and $2$. 
	\label{algorithm_DE_generate_mutant_vector}
	\State For each $i\in \{1,\dots,N_p\}$, introduce crossover to generate a trial vector $\mathbf{u}_i^l$ associated with $\mathbf{x}_i^l$. For $j\in\{1,\dots,D\}$, the $j$th element of $\mathbf{u}_i^l$ is given by
	\[
	u_{i,j}^l = \left\{
	\begin{array}{ll}
	v_{i,j}^l &  \mathrm{if}\ X[j]\leq \eta \ \mathrm{or}\ j = Y[i]\\
	x_{i,j}^l &  \mathrm{if}\ X[j]> \eta \ \mathrm{and}\ j \neq Y[i]
	\end{array}
	\right.
	\]
	where $X[j]$ are independent and identically distributed (i.i.d.) continuous random variables uniformly distributed in $(0,1)$, $Y[i]$ are i.i.d. discrete random variables uniformly distributed in $\{1,\dots,D\}$ and $\eta > 0$ is a constant threshold between $0$ and $1$.
	\label{algorithm_DE_generate_trial_vector}
	\State Using modified EXIT analysis, for each $i\in\{1,\dots,N_p\}$ find the iteration-constrained thresholds of $\mathbf{u}_i^l$ and $\mathbf{x}_i^l$. If $\mathbf{u}_i^l$ has the better threshold then set $\mathbf{x}_i^{l+1} = \mathbf{u}_i^l$, otherwise set $\mathbf{x}_i^{l+1} = \mathbf{x}_i^l$.
	\State Set $l = l+1 $ and repeat from step \ref{algorithm_DE_generate_mutant_vector} unless a stopping criterion has been reached.
\end{boxedalgorithmic}
\caption{Differential Evolution Algorithm}
\label{algorithm_DE}
\end{figure}

\section{Ensemble Design and Analysis}
\label{section_results}
\subsection{(G)LDPC Codes on the Binary Erasure Channel}
\label{section_BEC}

Using the approach outlined above, we obtained (G)LDPC code ensembles optimized for a constrained number of iterations on the BEC, with $R = 0.5$. The differential evolution process was initialized with a random population of cardinality $N_p = 70$ consisting of DDP vectors containing VNs of degrees $\{2, 3, \dots, 30\}$, and with generalized CNs consisting of degree-$7$ SPC codes, $(7,4)$ Hamming codes and $(15,11)$ Hamming codes. The optimization was performed for $i_{\mathrm{max}} = 10$ and the resulting ensemble was named Ensemble~A. Ensemble~B was generated by performing the same optimization, but with generalized CNs disallowed, i.e., with only degree-$7$ SPC codes at the CNs. Finally, Ensemble~C was optimized for $i_{\mathrm{max}} = 200$ iterations with generalized CNs allowed, as in the case of Ensemble~A. 

In order to compare the performance of codes from the ensembles, a single code having block length $N = 10,000$\footnote{In some cases a slight variation in the values of $N$ and $R$ is necessary in order to ensure an integer number of VNs and CNs while matching the DDP.} was constructed from each ensemble. These codes were generated by randomly assigning edge connections in accordance with the obtained VN and CN degree distributions, while ensuring no more than one edge connection between any VN and CN pair. By using long codes which are picked randomly from the ensembles in this way, we can have high confidence that the performances we observe are due to the intrinsic ensemble properties, rather than being a result of a particular algorithm being adopted in the construction of the finite-length code. For convenience, we refer to the codes constructed from Ensembles~A, B and C as Codes~A, B and C, respectively.

The DDPs for Ensembles~A, B and C are given in Table~\ref{table_Ensembles_A_B_C}, along with the iteration-constrained threshold, $\varepsilon^*$, obtained for each of these ensembles by our modified EXIT chart analysis.\footnote{In Ensemble~A, we note that although both $(7,4)$ and $(15,11)$ Hamming codes are allowed at the generalized CNs, the optimization has produced an ensemble which does not use any $(7,4)$ Hamming codes.}

\begin{table}[t]
\centering
\caption{Details for Ensembles A-C}
\label{table_Ensembles_A_B_C}
\begin{tabular}{| c | c | c | c |}
\cline{2-4}
\multicolumn{1}{c |}{} & Ensemble A & Ensemble B & Ensemble C \\
\hline
VN degree $d$ & \multicolumn{3}{ c |}{$\lambda_d$} \\
\hline
$2$ & $1$ &  & $0.318057$ \\
$3$ & & $0.841365$ & $0.202714$ \\
$4$ & & & $0.058171$ \\
$6$ & & & $0.147257$ \\
$13$ & & & $0.173086$ \\
$15$ & & & $0.100714$ \\
$30$ & & $0.158635$ & \\
\hline
CN type $t$ & \multicolumn{3}{ c |}{$\rho_t$} \\
\hline
$1$ & $0.134313$ & $1.000000$ & $1.000000$ \\
$2$ & $0.865687$ & & \\
\hline
$\varepsilon^*$ & $0.390459$ & $0.365436$ & $0.485836$ \\
\hline
$i_{\mathrm{max}}$ & $10$ & $10$ & $200$ \\
\hline
$\lambda^\prime(0)\rho^\prime(1)$ & -- & $0.000000$ & $1.908343$ \\
$\lambda^\prime(0)C$ & $0.805878$ & -- & -- \\
\hline
\multicolumn{4}{l}{} \\
\multicolumn{4}{l}{$t = 1$: Degree-$7$ SPC Code} \\
\multicolumn{4}{l}{$t = 2$: $(15,11)$ Hamming Code}
\end{tabular}
\end{table}

When optimized for a large number of iterations the fractions of edges connected to generalized CNs vanishes (Ensemble~C). In the iteration-constrained case, however, a significant fraction of the edges in the optimized ensemble are connected to generalized CNs when these CNs are allowed (Ensemble~A). In the case of Ensemble~C, the ensemble optimized for a large number of iterations, we also observe that the value of $\lambda_2$, the fraction of edges connected to degree-$2$ VNs is quite high. However for Ensemble~B, the LDPC ensemble optimized for a small number of iterations, $\lambda_2$ drops to zero.

\begin{figure}
\includegraphics[width=\linewidth]{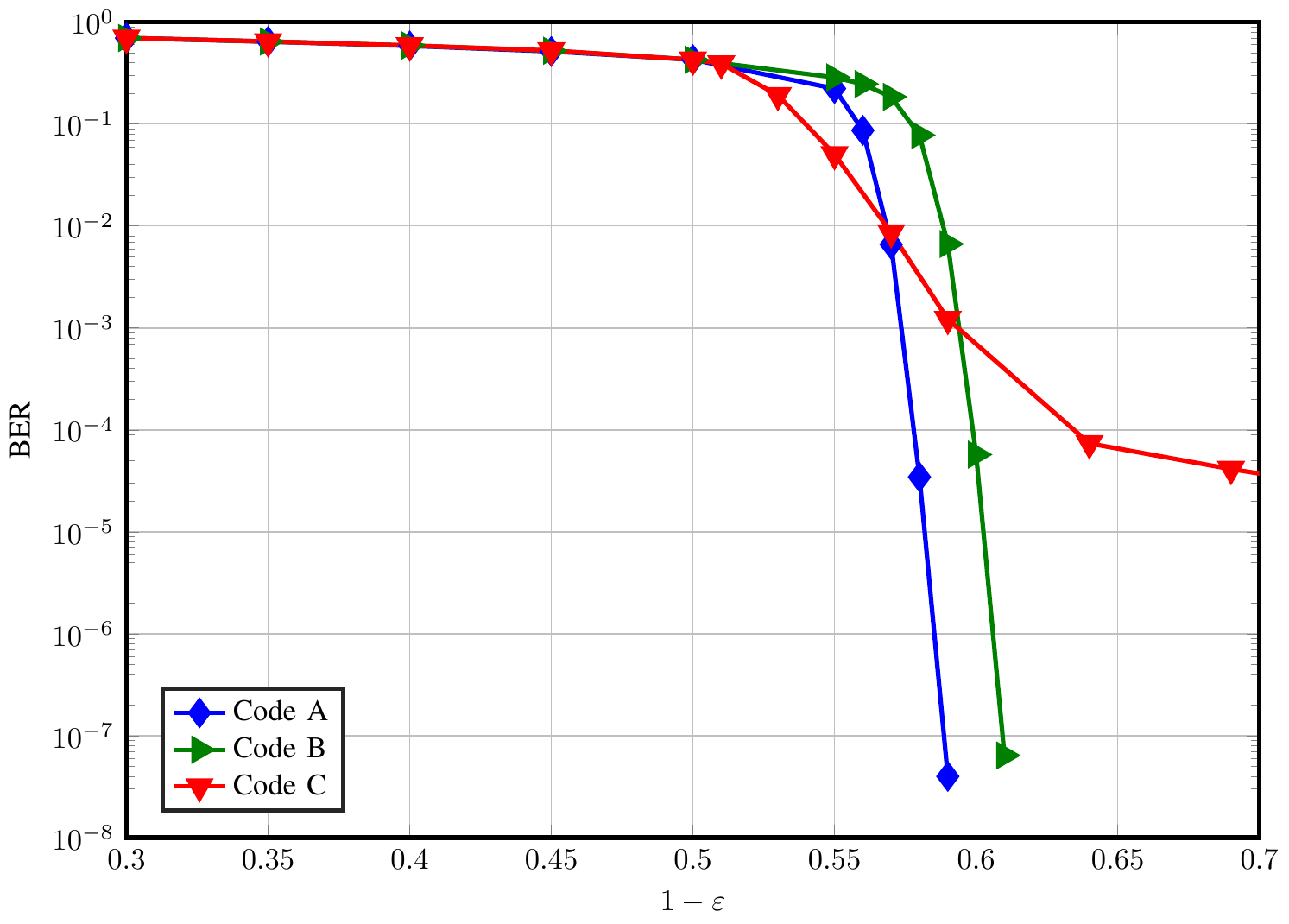}
\caption{BER of Codes~A, B and C over the BEC after $10$ decoder iterations. VN and CN degree distributions are given in Table~\ref{table_Ensembles_A_B_C}.}
\label{figure_BER_10_BEC}
\end{figure}

\begin{figure}
	\includegraphics[width=\linewidth]{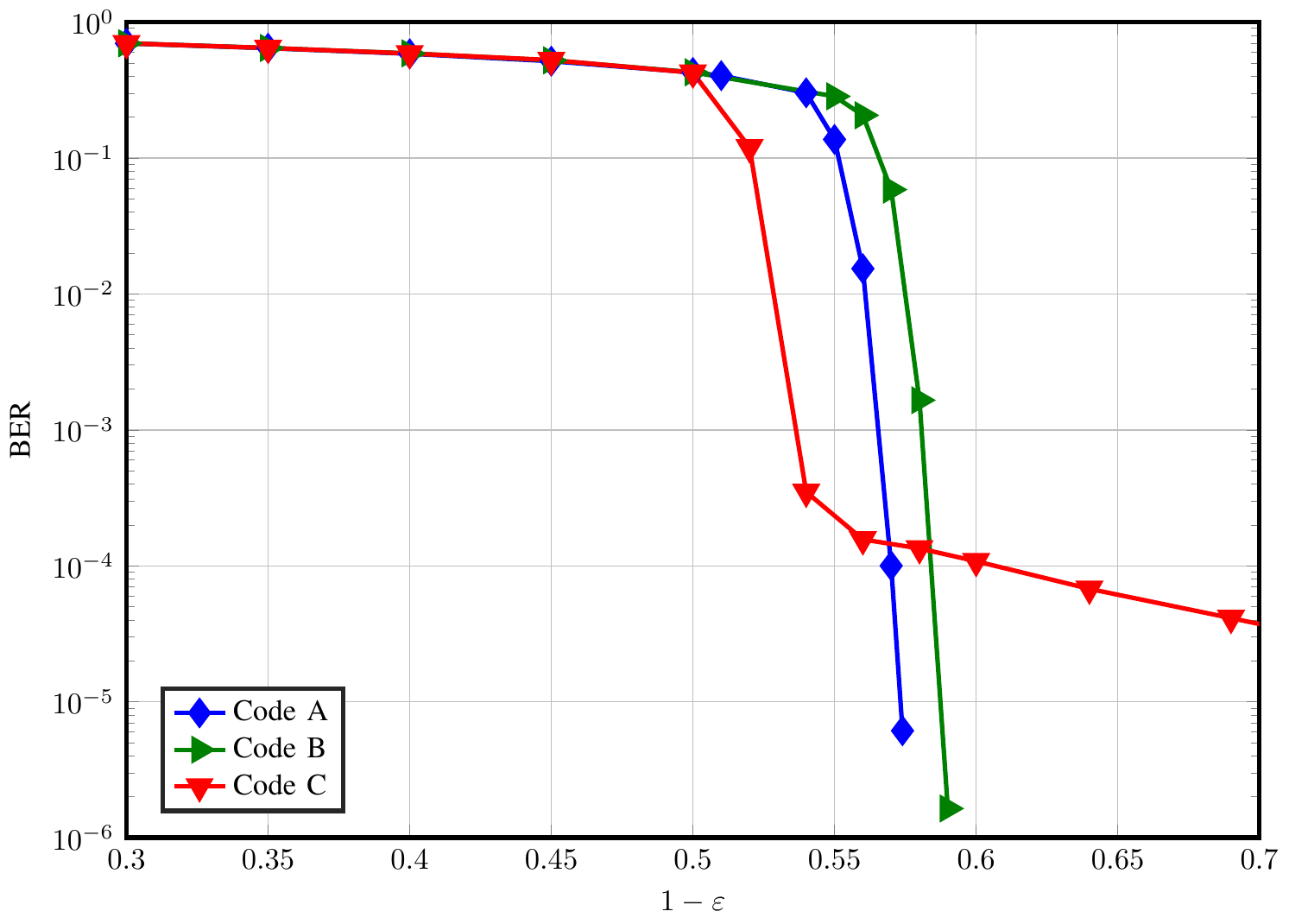}
	\caption{BER of Codes~A, B and C over the BEC after $200$ decoder iterations. VN and CN degree distributions are given in Table~\ref{table_Ensembles_A_B_C}.}
	\label{figure_BER_200_BEC}
\end{figure}

The bit error rate (BER) of Codes~A, B, and C after $10$ and $200$ belief propagation decoding iterations are shown in Fig. \ref{figure_BER_10_BEC} and Fig. \ref{figure_BER_200_BEC}, respectively. We focus on the BER instead of the codeword error rate (CER), as it is known that LDPC codes from ensembles with thresholds close to capacity (such as Code~C) offer excellent BER waterfall performance while their CER suffers due to poor distance properties. We are mainly interested in the waterfall regions of the finite-length codes. Significantly, we observe that when the decoder is limited to $i_{\max} = 10$ iterations, the codes from ensembles which are optimized specifically for $10$ iterations (Ensembles~A and B) still exhibit the very steep BER waterfall typical of iteratively decoded codes. Of these two codes, Code~A, the GLDPC code, outperforms its LDPC counterpart, Code~B, in accordance with their respective iteration-constrained thresholds $\varepsilon^*$. The poor performance of Code~C, the code from the ensemble optimized for $200$ iterations, in Fig. \ref{figure_BER_10_BEC} is imputable to the constrained number of iterations. Apart from its high error floor due to poor distance spectrum (which was expected, given the fact that $\lambda'(0)\rho'(1)$ is significantly larger than $1$), as Code~C has been designed for decoding within a maximum of $200$ iterations, the decoding path is typically far from complete when only $10$ iterations are allowed, resulting in the very low slope in the waterfall observed here. When $200$ decoder iterations are allowed, the waterfall for Code~C becomes very steep, showing a significant gain with respect to both Code~A and Code~B; however the poor minimum distance of Code~C prevents any significant improvement in the error floor.

As a first-order measure of complexity, it is significant to note that codes from Ensemble~B have fewer Tanner graph edges than codes of similar length from Ensemble~C. The GLDPC codes from Ensemble~A, in turn, have many fewer edges than LDPC codes of similar length from Ensemble~B. This, however, does not necessarily mean that codes from Ensemble~A will have a lower implementation complexity (or latency) as the processing at the generalized CNs of the GLDPC codes will be more complex than that of the SPCs at the CNs of the LDPC codes.

The iteration-constrained thresholds for Ensembles~A and B obtained by EXIT chart analysis are, as expected, lower than that of Code~C when the number of iterations is not constrained. In general, a code designed for $i_{\mathrm{max}}$ iterations will, like Code~C, perform poorly when fewer than $i_{\mathrm{max}}$ decoder operations are run. As such, the codes optimized for $i_{\max}$ iterations offer a ``middle-ground'' - having much larger achievable BEC erasure probabilities than Code~C for $i_\mathrm{max}$ iterations, but a much poorer threshold for unlimited iterations.

\subsection{LDPC Codes on the AWGN Channel}
\label{section_AWGNChannel}

Using the same procedure, LDPC code ensembles were also obtained for the AWGN channel. As in the BEC case, the DDPs in the starting population of the differential evolution algorithm contained VN degrees from $\{2,3,\dots,30\}$. The CN degrees in the initial DDPs ranged from $\{7, 8, \dots, 15\}$.

Using the same technique as before, codes were drawn from Ensemble~D, an ensemble optimized for an unlimited number of iterations, and Ensembles~E, F and G, which were optimized for $i_{\mathrm{max}} = 10, 20$ and $30$ iterations, respectively. Each of the codes have $R = 0.5$ and $N = 10,000$, as before. The DDPs and associated iteration-constrained thresholds, $(E_b/N_0)^*$, for these ensembles are given in Table~\ref{table_Ensembles_D-G}. Again, we refer to the codes constructed from Ensembles~D--G as Codes~D--G.

\begin{figure}
	\includegraphics[width=\linewidth]{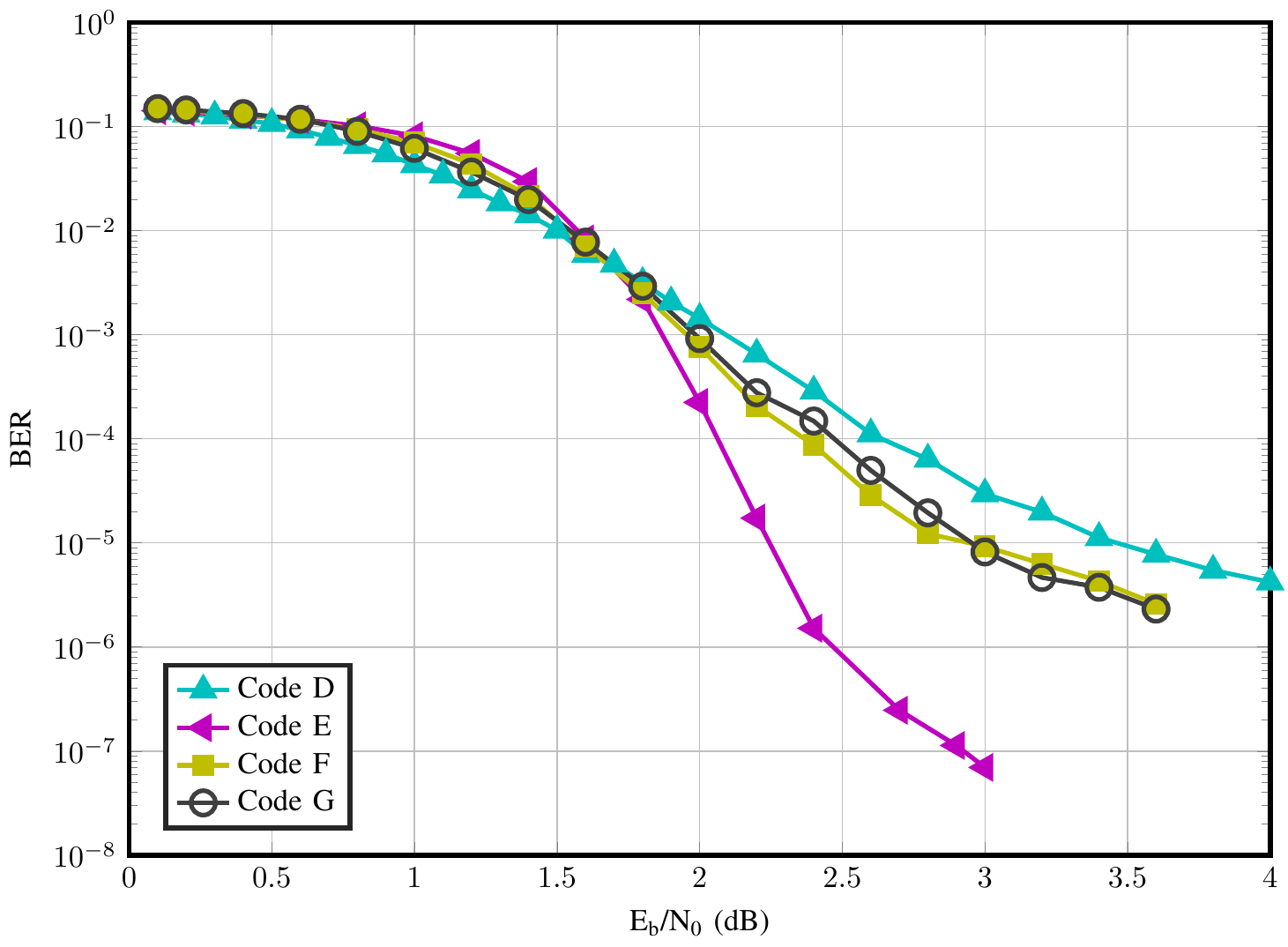}
	\caption{BER of Codes~D--G over the AWGN channel after $10$ decoder iterations. VN and CN degree distributions are given in Table~\ref{table_Ensembles_D-G}.}
	\label{figure_BER_10_AWGN}
\end{figure}

The BERs of these codes for $10$ decoding iterations are shown in Fig. \ref{figure_BER_10_AWGN}. Again, we observe that the most appealing performance for $i_{\mathrm{max}} = 10$ iterations is exhibited by the code that is designed for that particular number of iterations, and that the codes which have been designed for $i_{\mathrm{max}} > 10$ iterations exhibit poor performance when simulated for $10$ iterations. We note that these codes exhibit significantly higher error floors than Code~E (the code from the ensemble optimized for $10$ iterations). Similar to the BEC case, it is notable that Code~E has only a very small fraction of edges connected to degree-$2$ VNs, while in Codes~D, F and G, $\lambda_2$ contributes a significant fraction of the overall number of edges.

\begin{table}[t]
\centering
\caption{Details for Ensembles D--G}
\label{table_Ensembles_D-G}
\begin{tabular}{| c | c | c | c | c |}
\cline{2-5}
\multicolumn{1}{c |}{} & Ensemble & Ensemble & Ensemble & Ensemble\\
\multicolumn{1}{c |}{} & D & E & F & G\\
\hline
VN degree $d$ & \multicolumn{4}{ c |}{$\lambda_d$} \\
\hline
$2$ & $0.244010$ & $0.033563$ & $0.19128$ & $0.175711$ \\
$3$ & $0.154621$ & $0.567888$ & & $0.037810$ \\
$4$ & $0.065721$ & $0.068026$ & $0.307464$ & $0.279311$ \\
$5$ & $0.084352$ & & $0.061890$ & $0.068726$ \\
$6$ & $0.088753$ & & $0.083437$ & \\
$7$ & & & & $0.048109$ \\
$8$ & $0.039511$ & & $0.083481$ & \\
$12$ &  & $0.283606$ & & \\
$14$ &  & $0.046918$ & & \\
$18$ & $0.323032$ & & &\\
$30$ & & & $0.272447$ & $0.390333$ \\
\hline
CN degree $s$ & \multicolumn{4}{ c |}{$\rho_s$} \\
\hline
$7$ & & $0.001226$ & &\\
$8$ & $0.803716$ & $0.998775$ & &\\
$9$ & $0.196284$ & & $0.902024$ & $0.177118$\\
$10$ & & & & $0.822882$ \\
$11$ & & & $0.097976$ & \\
\hline
$(E_b/N_0)^*$ (dB) & $0.259614$ & $1.827213$ & $1.124776$ & $0.803599$ \\
\hline
$\lambda^\prime(0)\rho^\prime(1)$ & $3.660147$ & $0.436314$ & $3.443055$ & $2.987091$ \\
\hline
\end{tabular}
\end{table}

\subsection{Growth Rate and Distance Properties}
\label{Section_Analysis}

A useful tool in the analysis of LDPC and GLDPC code ensembles is the growth rate of their weight distribution (i.e., the asymptotic exponent of their weight distribution), given by 

\begin{equation*}
G(\alpha) = \lim_{N\rightarrow\infty}\frac{1}{N}\log\mathbb{E}_{\mathcal{M}_N}[A_{\alpha N}]
\end{equation*}

\noindent where $\alpha$ is the codeword weight normalized with respect to the block length $N$, $\mathbb{E}_{\mathcal{M}_N}$ is the expectation operator over the ensemble $\mathcal{M}_N$, and $A_w$ denotes the number of codewords of weight $w$ of a randomly chosen code in the ensemble $\mathcal{M}_N$ \cite{Di2006WeightDistributionLow, Flanagan2013SpectralShapeDoubly}.

An ensemble is said to have \emph{good growth rate behavior} if and only if the initial slope of the growth rate curve is negative. This is the case for LDPC and GLDPC codes when $\lambda^\prime(0)\rho^\prime(1) < 1$ or $\lambda^\prime(0)C < 1$, respectively \cite{Di2006WeightDistributionLow}. Also, it is desirable that the value $\alpha = \alpha^*$ at which $G(\alpha)$ crosses the horizontal axis is relatively large. Codes from ensembles which do not have good growth rate behavior typically exhibit a high error floor. From Tables \ref{table_Ensembles_A_B_C} and \ref{table_Ensembles_D-G} it may be seen that of Ensembles~A--G, only Ensembles~A, B and E have good growth rate behavior.

\begin{figure}[t]
	\includegraphics[width=\linewidth]{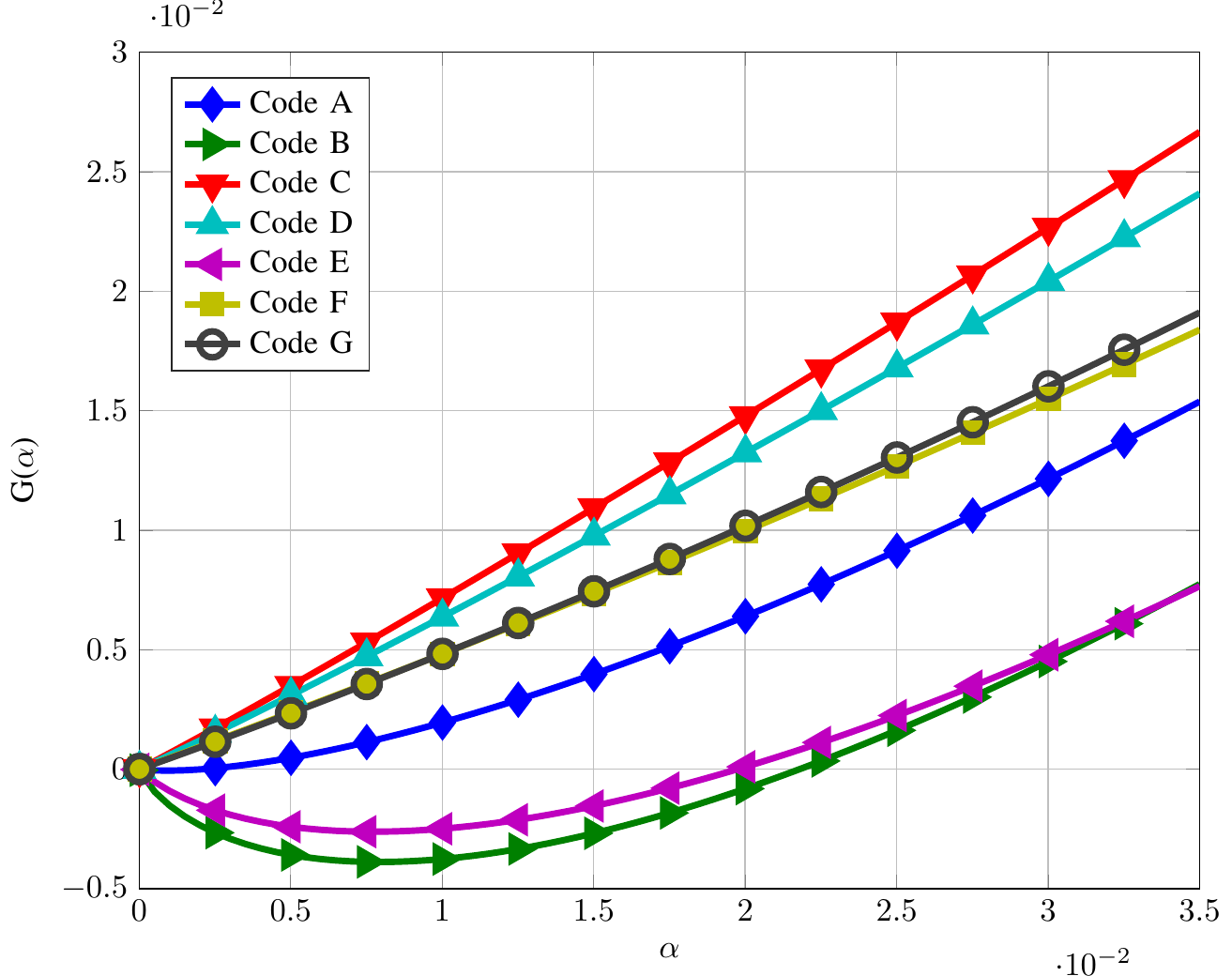}
	\caption{Growth rates of Ensembles~A--G. VN and CN degree distributions are given in Tables~\ref{table_Ensembles_A_B_C} and \ref{table_Ensembles_D-G}.}
	\label{figure_Growth_Rates}
\end{figure}

The growth rates of Ensembles~A--G are shown in Fig. \ref{figure_Growth_Rates}. As noted above, the poor growth rate behavior of Ensembles~C, D, F and G leads us to expect that codes drawn from these ensembles will have high error floors, as seen in Fig. \ref{figure_BER_10_BEC} and Fig. \ref{figure_BER_10_AWGN}. The poor performance of these codes after $10$ iterations is therefore a consequence of both their inability to complete their long decoding paths within a small number of iterations and the poor growth rate behavior of their ensembles. Indeed, as seen in the case of Code~C in Fig. \ref{figure_BER_200_BEC}, even when simulated for the number of iterations for which they are optimized, these codes exhibit high error floors as a result of their bad growth rate behavior.

In order to empirically verify that LDPC codes designed for a small number of decoding iterations are very favorable in terms of minimum distance (which is suggested by the results obtained for the asymptotic growth rate of the weight distribution), we performed the following experiment. A constrained ensemble optimization based on EXIT charts and differential evolution was carried out for an infinite number of iterations, but imposing the constraint $\lambda^\prime(0)\rho^\prime(1)<0.5$ in order to enforce a good growth rate behavior. This is a classical ensemble design strategy to obtain a good trade-off between waterfall and error floor performance. The obtained DDP is $\lambda(x)=0.062498\,x + 0.479743\,x^2 + 0.049808\,x^5 + 0.117758\,x^8 + 0.290192\,x^{29}$ and $\rho(x)=x^8$. Next, we constructed a finite length $(1024,512)$ LDPC code according to this DDP using the progressive edge growth (PEG) algorithm \cite{Hu2005Regularandirregular} and analyzed its weight distribution through the algorithm proposed in \cite{Hu2004computationminimumdistance}. The minimum distance of this code was found to be $44$. We then constructed via the PEG algorithm a $(1024,512)$ code from Ensemble~E and analyzed its weight distribution via the same algorithm. The minimum distance of this second code was found to be substantially larger, in particular equal to $58$. This strengthens our conclusion that very good minimum distance properties are inherent to LDPC codes designed for iterative decoders severely constrained in terms of maximum number of iterations.

\bibliography{IEEEabrv.bib,main.bib}{}
\bibliographystyle{IEEEtran}

\end{document}